\documentclass[onecolumn,showpacs,preprintnumbers]{revtex4}
\usepackage{mathrsfs}
\usepackage{graphicx}
\usepackage{dcolumn}
\usepackage{bm}
\usepackage{epsfig}
\usepackage{amsmath,amssymb}
\usepackage{multirow}
\usepackage{CJK}

\begin{document}
\title{The analysis of the excited bottom and bottom strange states $B_{1}(5721)$, $B_{2}^{*}(5747)$,
$B_{s1}(5830)$, $B_{s2}^{*}(5840)$, $B_{J}(5840)$ and $B_{J}(5970)$ in B meson family.}
\author{Guo-Liang YU}
\email{yuguoliang2011@163.com}
\author{Zhi-Gang WANG}
\email{zgwang@aliyun.com}
\affiliation{ Department of Mathematics and Physics, North China
Electric power university, Baoding 071003, People's Republic of
China}
\date{\today }

\thanks{supported by the Fundamental Research Funds for
the Central Universities, Grant Number $2016MS133$ and Natural Science Foundation of HeBei Province, Grant Number
A2018502124.}

\begin{abstract}
In order to make a further confirmation about the assignments of the excited bottom and bottom strange mesons $B_{1}(5721)$, $B_{2}^{*}(5747)$,
$B_{s1}(5830)$, $B_{s2}^{*}(5840)$ and meanwhile identify the possible assignments of $B_{J}(5840)$, $B_{J}(5970)$, we study the strong decays
of these states with the $^{3}P_{0}$ decay model. Our analysis support $B_{1}(5721)$ and $B_{2}^{*}(5747)$ to be the $1P_{1}'$ and $1^{3}P_{2}$
assignments and the $B_{s1}(5830)$, $B_{s2}^{*}(5840)$ to be the strange partner of $B_{1}(5721)$ and $B_{2}^{*}(5747)$. Besides, we tentatively
identify the recently observed $B_{J}(5840)$, $B_{J}(5970)$ as the $2^{3}S_{1}$ and $1^{3}D_{3}$ states, respectively. It is noticed that this conclusion
needs further confirmation by measuring the decay channel to $B\pi$ of $B_{J}(5840)$ and $B_{J}(5970)$ in experiments.
\\
\textbf{Key words:} Bottom mesons, $^{3}P_{0}$ model, Strong decays
\\
\textbf{PACS:} 13.25.Ft, 14.40.Lb
\end{abstract}
\maketitle

\begin{Large}
\textbf{1 Introduction}
\end{Large}

In recent decades, theoretical and experimental physicists have made a progress in studying the
heavy-light meson spectrum with the observation of a large number of charmed and bottom mesons.
Especially, the charmed meson spectrum has been mapped out with high precision with the observation of many new
charmed states such as $D_{1}^{*}(2680)$,
$D_{2}^{*}(2460)$, $D_{J}(2580)$, $D_{J}^{*}(2650)$, $D_{J}^{*}(2760)$, $D_{J}(2740)$, $D_{J}(3000)$, $D_{J}^{*}(3000)$, etc.\cite{Aaij1,Aaij2,Del1}.
In our previous work, we studied the
strong decay behaviors of some charmed states with the $^{3}P_{0}$ decay model and the heavy meson effective theory, and identified the quantum numbers
of these charmed states\cite{Guo1,Wang1,Wang2}.
Whereas for bottom sector, only the ground states, $B^{0}(5279)$, $B^{\pm}(5279)$, $B^{*}(5324)$, $B_{s}(5366)$,
$B_{s}^{*}(5415)$ and a few of low lying excited states, $B_{1}(5721)$, $B_{2}^{*}(5747)$ have been identified
in PDG\cite{PDG}. Comparing with the charmed mesons, we know little about the information of most of the excited bottom states

Fortunately for us, the LHCb collaboration have observed some new bottom states in recent years, such as $B_{J}(5721)^{0}$, $B_{J}(5721)^{+}$,
$B_{2}^{*}(5747)^{0}$, $B_{2}^{*}(5747)^{+}$, $B_{J}(5840)^{0}$, $B_{J}(5840)^{+}$, $B_{J}(5970)^{0}$, $B_{J}(5970)^{+}$\cite{B1,B2,B3,B7}.
Besides, CDF, D0 and LHCb collaborations have also observed two bottom strange mesons, $B_{s1}(5830)$, $B_{s2}^{*}(5840)$\cite{B4,B5,B6} and
assigned its $J^{P}$ to be $1^{+}$ and $2^{+}$, respectively. The masses and the widths of these newly observed bottom and bottom strange
mesons are listed in Table I. For these mesons, an important work is to identify its quantum numbers and assign a place in the
bottom meson spectrum. We can adopt several approaches to carry out this work such as quark model\cite{M1,M2,M3}, Heavy Quark Effective Theory(HQET)\cite{M4,Wang1},
lattice QCD\cite{Lattice1} and $^{3}P_{0}$ model\cite{3P01,3P02,3P03} etc. However, the predictions obtaining from different theoretical approaches,
even the same theoretical method with different parameters are not completely consistent with each other.
\begin{table*}[htbp]
\begin{ruledtabular}\caption{The experimental information about the excited bottom and bottom strange states in this paper.}
\begin{tabular}{c c c c c c }
States & \  Mass(MeV/c$^{2}$)  & \ Width(MeV)  & \ $J^{PC}$ &\ Decay channels  \\
\hline
$B_{1}(5721)^{+}$ \cite{PDG} & \  $5725.9^{+2.5}_{-2.7}MeV$         &  \ $31\pm6MeV$     & \  $1^{+}$   &  \   $B^{*0}\pi^{+}$   \\
$B_{1}(5721)^{0}$\cite{PDG} & \   $5726.1\pm1.3MeV$         &  \ $27.5\pm3.4MeV$     & \  $1^{+}$    &  \   $B^{*+}\pi^{-}$  \\
$B_{2}^{*}(5747)^{+}$\cite{PDG} & \   $5737.2\pm0.7MeV$       &  \ $20\pm5$    & \  $2^{+}(1^{3}P_{2})$    &  \  $B^{0}\pi^{+}$, $B^{*0}\pi^{+}$   \\
$B_{2}^{*}(5747)^{0}$\cite{PDG}        &\    $5739.5\pm0.7$  &\   $24.2\pm1.7$ &\  $2^{+}(1^{3}P_{2})$  &\ $B^{+}\pi^{-}$, $B^{*+}\pi^{-}$ \\
$B_{J}(5970)^{+}$\cite{PDG}    & \   $5964\pm5MeV$         &  \  $ 62\pm20$      & \   -   &  \   $B^{*0}\pi^{+}$, [$B^{0}\pi^{+}$]    \\
$B_{J}(5970)^{0}$\cite{PDG}       & \   $5971\pm5MeV$\cite{PDG}        &  \  $81\pm12$   & \   -   &  \   $B^{*0}\pi^{+}$, [$B^{+}\pi^{-}$] \\
$B_{J}(5840)^{+}$\cite{B7}  & \   $5862.9\pm5.0$       &  \  $224\pm80MeV$   & \    -   &  \  $B^{*0}\pi^{+}$, [$B^{0}\pi^{+}$]  \\
$B_{J}(5840)^{0}$\cite{B7}  & \   $5862.9\pm5.0$        &  \  $127.4\pm16.7MeV$   & \    -   &  \  $B^{*0}\pi^{+}$, [$B^{+}\pi^{-}$]  \\
$B_{s1}(5830)$\cite{PDG}  & \   $5828.7\pm0.1$       &  \  $0.5\pm0.3$    & \   $1^{+}$   &  \   $B^{*}K$   \\
$B_{s2}^{*}(5840)$\cite{PDG}      & \   $5839.85\pm0.7$        &  \   $1.40\pm0.4$   & \  $2^{+}(1^{3}P_{2})$    &  \   $B^{*}K$,$BK$   \\
\end{tabular}
\end{ruledtabular}
\end{table*}

Since the discoveries of the bottom mesons $B_{1}(5721)$ and $B_{2}^{*}(5747)$ by the $D0$ collaboration in 2007\cite{B1},
people studied its nature with different models and identified these two mesons as the $1^{+}$ and $2^{+}$ bottom states
in PDG\cite{PDG}. However, it is still need confirmation for the assignments of the $B_{1}(5721)$ meson because it is the mixing
of the $^{3}P_{1}$ and $^{1}P_{1}$ states. For $B_{J}(5970)$ bottom meson, it was mainly explained to be a $2S1^{-}$ or
$1D3^{-}$ state by different theoretical approaches\cite{701,702,703,704,705,706,707}. And its spin parity still remain undetermined in the PDG, which
only listed its mass and decay width. Further more, we note that the $B_{J}(5840)$ meson was omitted from the summary tables
in the PDG, which indicates that the assignment of this meson needs more theoretical and experimental verifications. As for the $B_{s2}^{*}(5840)$
and $B_{s1}(5830)$ bottom-strange mesons, people assigned these two mesons as the strange parters of $B_{2}^{*}(5747)$ and $B_{1}(5721)$
with quantum numbers to be $2^{+}$ and $1^{+}$ respectively\cite{PDG,301,302,303,304}.

In our previous work, we studied the two-body strong decays of the $B_{1}(5721)$, $B_{2}^{*}(5747)$, $B(5970)$, $B_{s1}(5830)$ and $B_{s2}(5840)$
with the heavy meson effective theory in the leading order approximation, and assigned states $2S1^{-}$, $1D1^{-}$
and $1D3^{-}$ as the candidate of $B_{J}(5970)$\cite{701}. As a continuation of our previous work, we study the strong decay behaviors of more
bottom mesons with the $^{3}P_{0}$ decay model and give a simple discussion about the quantum numbers of these mesons.
The calculated strong decay widths in this work will be confronted with the experimental data in the future and will be helpful in determining the nature
of these heavy-light mesons.  This article is arranged as follows: In section 2, we give a brief review
of the $^{3}P_{0}$ decay model; in Sec.3 we study the strong decays
of $B_{1}(5721)$, $B_{2}^{*}(5747)$,
$B_{s1}(5830)$, $B_{s2}^{*}(5840)$, $B_{J}(5840)$ and $B_{J}(5970)$ and identify the assignments of these states; in
Sec.4, we present our conclusions.

\begin{Large}
\textbf{2 Strong decay model}
\end{Large}
\begin{figure}[h]
  \includegraphics[width=15cm]{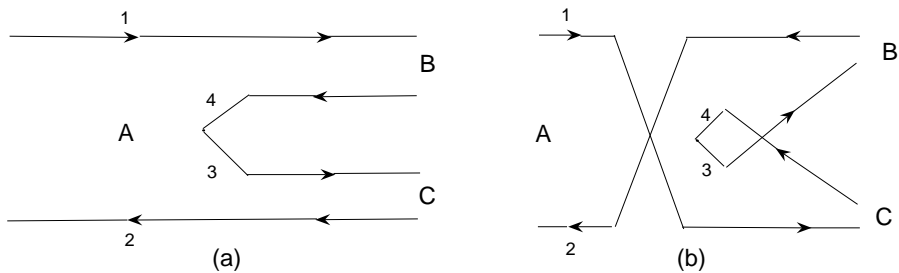}
  \caption{The two possible diagrams contributing to $A\rightarrow BC$ in the $^{3}P_{0}$ model.}\label{Figure 1:}
\end{figure}

To study the strong decay properties of
the mesons, the $^{3}P_{0}$ decay model is an effective and simple method, which
can give a good prediction about the decay behaviors of many
hadrons\cite{3P1,3P2,3P3,3P4,3P5}.
This model was first introduced by Micu in 1969\cite{3P01} and further developed
by Le Yaouanc and other collaborations\cite{3P02,3P03}. In Ref.\cite{Barnes} Barnes et al. performed a comprehensive study of light meson strong decays with
the $^{3}P_{0}$ model. Now, this model has
been extensively used to describe the strong decays of the heavy
mesons in the charmonium\cite{3P6,3P7,3P8,3P9} and
bottommonium systems\cite{3P10,3P11,3P12}, the
baryons\cite{3P13} and even the teraquark states\cite{3P14}.

At first, people considered an alternative phenomenological model to study the strong decays, in which
quark-antiquark pairs are produced with $^{3}S_{1}$ quantum numbers. However, this possibility is disfavoured by measuring ratios
of partial wave amplitudes\cite{Geiger}. In $^{3}P_{0}$ decay model, it is now generally accepted that a quark-antiquark pair($q_{3}\overline{q}_{4}$)
with $0^{++}$ quantum numbers(in the $^{3}P_{0}$ state) is
created from the vacuum\cite{3P01,3P02,3P03,3P1}. For a
meson decay process $A$$\rightarrow$$BC$, the quark-antiquark pair($q_{3}\overline{q}_{4}$) regroups into final state mesons($BC$) with the
$q_{1}\overline{q}_{2}$ from the initial meson $A$. This process is illustrated in FIG.1 and its transition operator in
the nonrelativistic limit is written as,
\begin{equation}
\begin{split}
    T=
    &-3\gamma\sum_{m}\langle1m1-m\mid00\rangle\int d^{3}\vec p_{3}d^{3}\vec p_{4}\delta^{3}(\vec p_{3}+\vec p_{4})\mathcal{Y}_{1}^{m}(\frac{\vec p_{3}-\vec p_{4}}{2})
    \chi_{1-m}^{34}\varphi_{0}^{34}\omega_{0}^{34}q_{3}^{\dag}(\vec p_{3})q_{4}^{\dag}(\vec p_{4})
\end{split}
\end{equation}
where $q_{3}^{\dag}$ and $q_{4}^{\dag}$ are the creation operators
in the momentum-space for the quark-antiquark
$q_{3}\overline{q}_{4}$ pair. $\gamma$ is a dimensionless parameter
reflecting the creation strength of the quark-antiquark pair. $\varphi_{0}^{34}$, $\omega_{0}^{34}$ and $\chi_{1-m}^{34}$
denote its flavor, color and spin wave functions.

In the c.m. frame, the amplitude of a decay process $A\rightarrow BC$ can be written as,
\begin{equation}
\begin{aligned}
\mathcal{M}^{M_{J_{A}}M_{J_{B}}M_{J_{C}}}(\vec P)=
&\gamma\sqrt{8E_{A}E_{B}E_{C}}\sum_{\mbox{\tiny$\begin{array}{c}
M_{L_{A}},M_{S_{A}},\\
M_{L_{B}},M_{S_{B}},\\
M_{L_{C}},M_{S_{C}},m\end{array}$}}\langle
L_{A}M_{L_{A}}S_{A}M_{S_{A}}\mid J_{A}M_{J_{A}}\rangle \langle
L_{B}M_{L_{B}}S_{B}M_{S_{B}}\mid J_{B}M_{J_{B}}\rangle \\
&\times\langle L_{C}M_{L_{C}}S_{C}M_{S_{C}}\mid
J_{C}M_{J_{C}}\rangle\langle 1m1-m\mid 00\rangle\langle \chi_{S_{B}M_{S_{B}}}^{14}\chi_{S_{C}M_{S_{C}}}^{32}\mid \chi_{S_{A}M_{S_{A}}}^{12}\chi_{1-m}^{34}\rangle \\
&\times \Big[\langle \phi_{B}^{14}\phi_{C}^{32}\mid \phi_{A}^{12}\phi_{0}^{34}\rangle I(\vec P,m_{1},m_{2},m_{3}) \\
&+(-1)^{1+S_{A}+S_{B}+S_{C}}\langle \phi_{B}^{32}\phi_{C}^{14}\mid
\phi_{A}^{12}\phi_{0}^{34}\rangle I(-\vec P,m_{2},m_{1},m_{3})\Big],
\end{aligned}
\end{equation}
where $\langle\chi_{S_{B}M_{S_{B}}}^{14}\chi_{S_{C}M_{S_{C}}}^{32}\mid \chi_{S_{A}M_{S_{A}}}^{12}\chi_{1-m}^{34}\rangle$, $\langle \phi_{B}^{14}\phi_{C}^{32}\mid \phi_{A}^{12}\phi_{0}^{34}\rangle$
are the spin and flavor matrix elements. The two terms in the last factor correspond to the two possible diagrams in FIG.1. The momentum space integral $I(\vec
P,m_{1},m_{2},m_{3})$ is given by
\begin{equation}
\begin{split}
I(\vec P,m_{1},m_{2},m_{3})=
&\int d^{3}\vec p \psi^{*}_{n_{B}L_{B}M_{L_{B}}}(\frac{m_{3}}{m_{1}+m_{2}}\vec P_{B}+\vec p)\psi^{*}_{n_{C}L_{C}M_{L_{C}}}(\frac{m_{3}}{m_{2}+m_{3}}\vec P_{B}+\vec p) \\
&\times\psi_{n_{A}L_{A}M_{L_{A}}}(\vec P_{B}+\vec p)\mathcal{Y}_{1}^{m}(\vec p)
\end{split}
\end{equation}
where $\vec P =\vec P_{B} =-\vec P_{C}, \vec p = \vec p_{3}$,
$m_{3}$ is the mass of the created quark $q_{3}$. In Eq.(3), $\psi$ is the simple harmonic oscillator (SHO) function which is use to describe the space part of the meson.
In momentum space, it is defined as
\begin{equation}
\begin{split}
\Psi_{nLM_{L}}(\vec p)=
&(-1)^{n}(-i)^{L}R^{L+\frac{3}{2}}\sqrt{\frac{2n!}{\Gamma(n+L+\frac{3}{2})}}exp(-\frac{R^{2}p^{2}}{2})L_{n}^{L+\frac{1}{2}}(R^{2}p^{2})\mathcal{Y}_{LM_{L}}(\vec
p)
\end{split}
\end{equation}
where $R$ is the scale parameter of the SHO. With the Jacob-Wick
formula, we can convert the helicity amplitude into the partial
wave amplitude
\begin{equation}
\begin{split}
\mathcal{M}^{JL}(\vec P)=
&\frac{\sqrt{4\pi(2L+1)}}{2J_{A}+1}\sum_{M_{J_{B}}M_{J_{C}}}\langle
L0JM_{J_{A}}|J_{A}M_{J_{A}}\rangle \langle
J_{B}M_{J_{B}}J_{C}M_{J_{C}}|JM_{J_{A}}\rangle\mathcal{M}^{M_{J_{A}}M_{J_{B}}M_{J_{C}}}(\vec
P)
\end{split}
\end{equation}
where $M_{J_{A}}=M_{J_{B}}+M_{J_{C}}$, $\mathbf{J_{A}=J_{B}+J_{C}}$
and $\mathbf{J_{A}+J_{P}=J_{B}+J_{C}+L}$.

The decay width in terms of partial wave amplitudes using the relative phase space is
\begin{equation}
\Gamma=\frac{\pi}{4}\frac{|\vec P|}{M_{A}^{2}}\sum_{JL}|\mathcal{M}^{JL}|^{2}
\end{equation}
where $P=|\vec
P|=\frac{\sqrt{[M_{A}^{2}-(M_{B}+M_{C})^{2}][M_{A}^{2}-(M_{B}-M_{C})^{2}]}}{2M_{A}}$ is the three momentum of the daughter mesons in the c.m. frame.
$M_{A}$, $M_{B}$, and $M_{C}$ are the masses of the mesons $A$, $B$,
and $C$, respectively. One can consult references\cite{3P01,3P02,3P03,3P1} for more details of the decay model.

\begin{Large}
\textbf{3 The results and discussions}
\end{Large}

The parameters involved in the $^{3}P_{0}$ model include the light quark pair($q\overline{q}$) creation
strength $\gamma$, the SHO wave function scale parameter $R$, and
the masses of the mesons and the constituent quarks. First, the masses of the quark are taken as $m_{u} = m_{d} =
0.22$ GeV, $m_{s} = 0.42$ GeV and $m_{b} = 4.81GeV$\cite{PDG}. Second, as for the factor $\gamma$, it describes the strength of quark-antiquark
pair creation from the vacuum and its value needs to be fitted according to experimental data. We take the fitted
value $\gamma=6.25$ for $u/d$ quark and $\gamma_{s\overline{s}}=\gamma/\sqrt{3}$ for $s$ quark\cite{3P1}.
This value is higher than that used by Kokoski and Isgur by a factor of $\sqrt{96\pi}$ due to different field
theory conventions, constant factors in $T$, etc\cite{Kokoski}.

The input parameter $R$ has a significant influence on the
shape of the radial wave function, which lead to the spatial integral of Eq.(3) being sensitive to the parameter
$R$. Thus, the decay width based on the $^{3}P_{0}$ decay model is also sensitive to the parameter $R$. Taking the strong decay of $B_{2}^{*}(5747)$ as an example, we plot
the decay width versus the input parameters $R$ in FIG.2. We can clearly see the dependence of the decay widths on the input parameter $R$. If the $R_{B^{0}}$, $R_{B^{+}}$,
$R_{B^{*0}}$, $R_{B^{*+}}$ and $R_{\pi}$ are fixed to be $2.5GeV^{-1}$, the decay width of $B_{2}^{*}(5747)$ changes several times with the value of $R_{B_{2}^{*}(5747)}$
changing from $2.0GeV^{-1}$ to $3.0GeV^{-1}$. As for this problem, there are two kinds of choices which are the common value and the effective value. The effective value
is fixed to reproduce the realistic root mean square radius by solving the Schrodinger equation with a linear potential. For the common value, H.G. Blunder et al\cite{3P1} carry out a
series of least squares fits of the model predictions to the decay widths of $28$ of the best known meson decays. And the common oscillator parameter $R$ with the vaue $2.5GeV^{-1}$
is suggested to be optimal. In our previous work, we studied strong decays of some charmed mesons with common value and obtained consistent results with experimental data. Thus, we still adopt
common value as the input of $R$ in this work.

\begin{figure}[h]
\begin{minipage}[t]{0.45\linewidth}
\centering
\includegraphics[height=5cm,width=7cm]{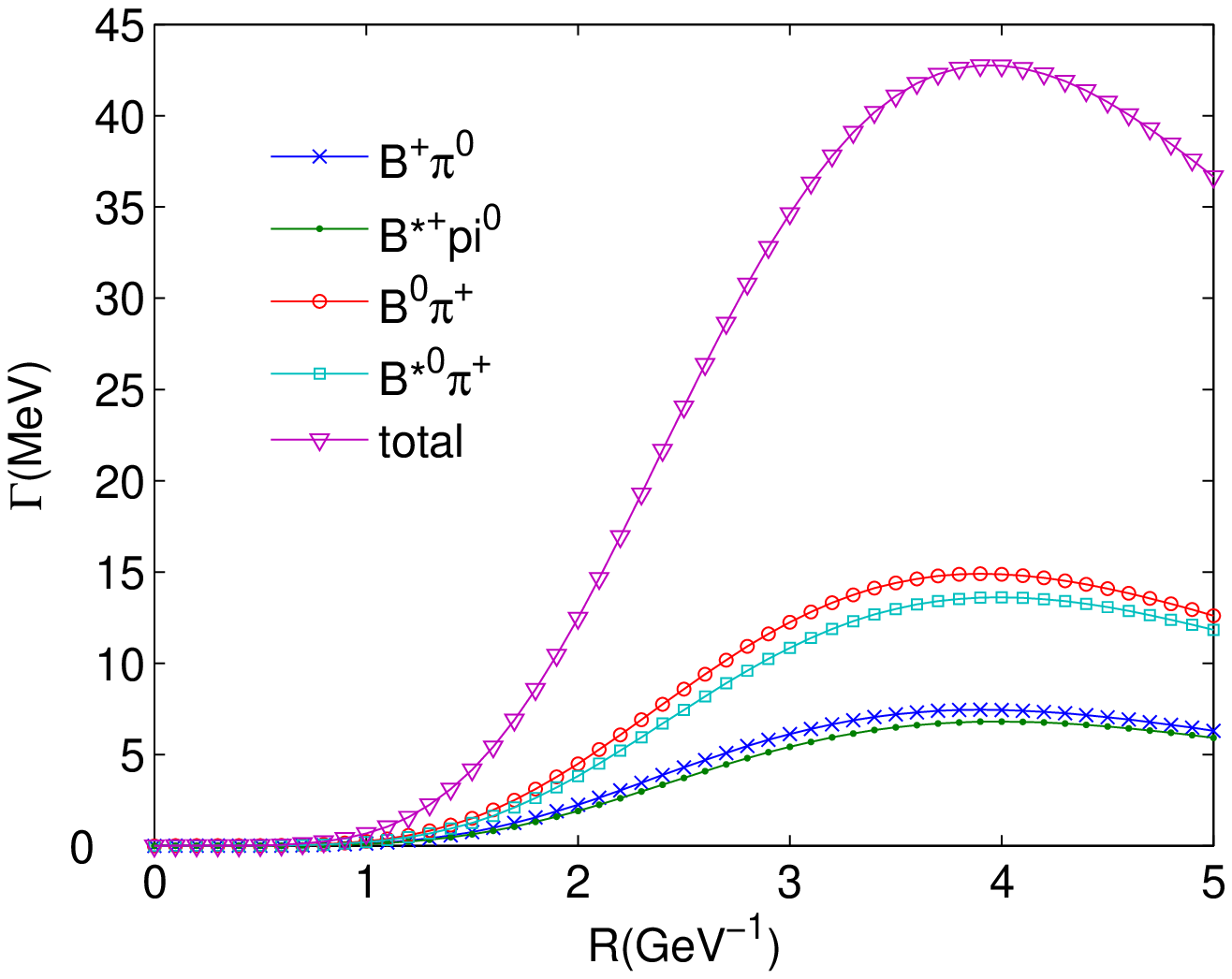}
\caption{The strong decay of $B^{*+}_{2}(5747)$ as the
$1^{3}P_{2}$ state on scale parameter
$R$.\label{your label}}
\end{minipage}
\hfill
\begin{minipage}[t]{0.45\linewidth}
\centering
\includegraphics[height=5cm,width=7cm]{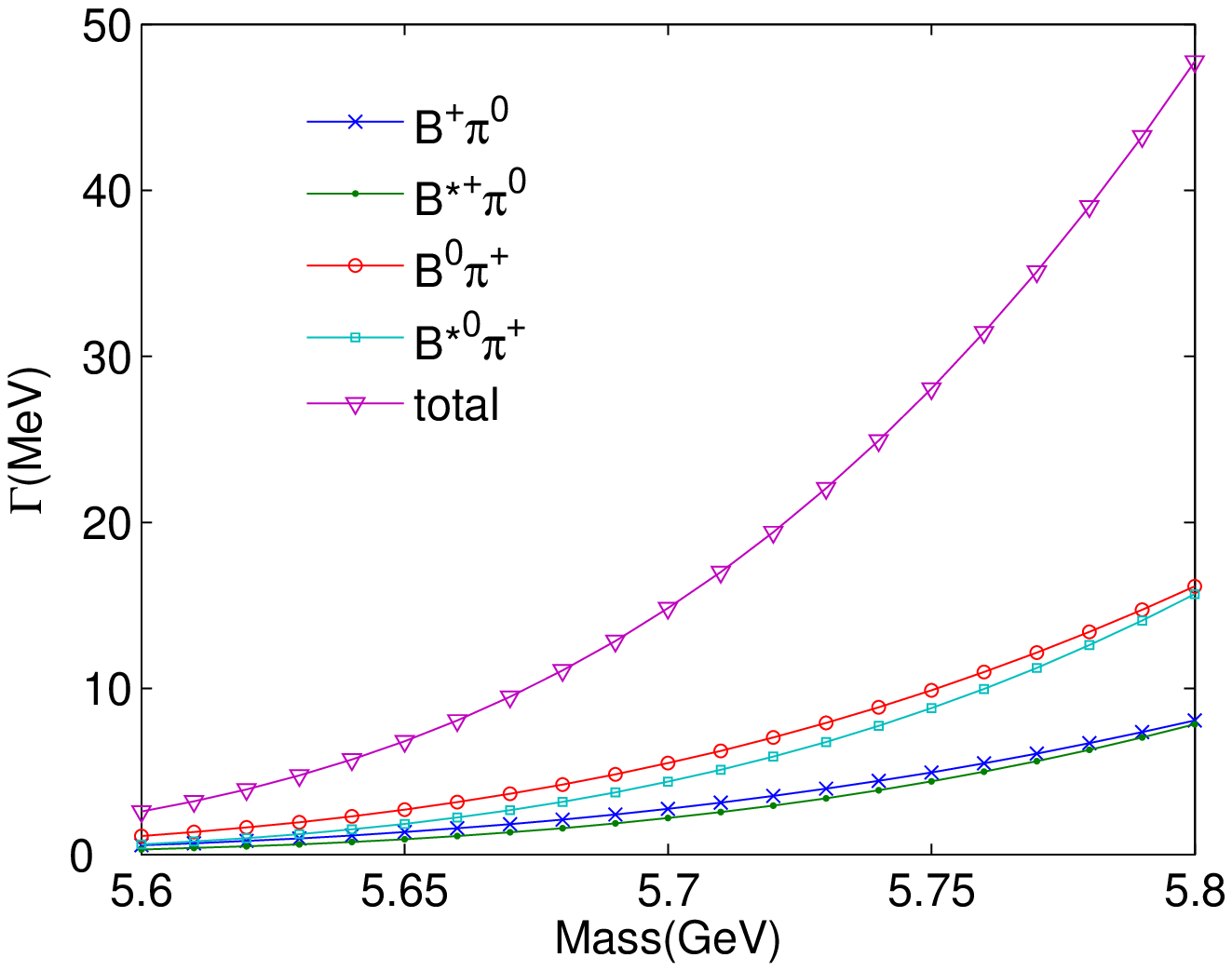}
\caption{The strong decay of $B^{*+}_{2}(5747)$ as the
$1^{3}P_{2}$ state on the mass.\label{your label}}
\end{minipage}
\end{figure}
\begin{table*}[htbp]
\begin{ruledtabular}\caption{The adopted masses of the hadrons used in our calculations.}
\begin{tabular}{c c c c c c c c c c c c c c c c c c c c c}
States & \  $\pi^{\pm}$  & \ $\pi^{0}$  & \ $\eta$ &\ $B^{\pm}$ & \ $B^{0}$   \\
Mass(MeV) & \ $139.6$         &  \ $135.0$     & \  $547.9$    &  \   $5279.3$   & \  5279.6     \\
\hline States  &\ $B^{*}$ & \ $B_{s}^{0}$ & \ $B_{s}^{*}$  &\ $K^{\pm}$  &\ $\overline{K^{0}}$  \\
 Mass(MeV)  &\ $5324.7$ &  \  $5366.9$    &  \  $5415.4$ & \ $493.67$ &\  $497.61$   \\
\end{tabular}
\end{ruledtabular}
\end{table*}

Finally, the mass of the meson has also a significant influence on the strong decay of the studied meson. For $B_{2}^{*}(5747)$ as an example,
if the masses of the daughter mesons are taken to be the standard values in PDG, the decay widths of $B_{2}^{*}(5747)$ vary greatly with its mass, which
can be seen in FIG.3. We know that the masses of the bottom mesons, especially the newly observed bottom states, have been updated from time to time.
In this work, we take the recently updated values in PDG\cite{PDG} as the input and list these values in TABLE II. As for the newly
observed bottom states which were omitted in PDG, we take the experimental data as the input.

It is noticed that mixing can occur between states with $J=L$ and $S=1$ or $S=0$. The relation between the heavy quark symmetric states and
the non-relativistic states $^{3}L_{L}$ and $^{1}L_{L}$ is written as\cite{Matsuki},

\begin{equation}
{
\left( \begin{array}{ccc}
|s_{l}=L+\frac{1}{2},L^{P}\rangle \\
|s_{l}=L-\frac{1}{2},L^{P}\rangle
\end{array}
\right )}=\frac{1}{\sqrt{2L+1}}{
\left( \begin{array}{ccc}
\sqrt{L+1} & -\sqrt{L} \\
\sqrt{L} & \sqrt{L+1}
\end{array}
\right )}{
\left( \begin{array}{ccc}
|^{3}L_{L} \\
|^{1}L_{L}
\end{array}
\right )}
\end{equation}
For the states $J=L=1$, the mixture angle is $\theta=-54.7^{\circ}$ or $\theta=35.3^{\circ}$, thus this relation transforms into
\begin{equation}
{
\left( \begin{array}{ccc}
|\frac{3}{2},1^{+}\rangle \\
|\frac{1}{2},1^{+}\rangle
\end{array}
\right )}={
\left( \begin{array}{ccc}
cos\theta & -sin\theta \\
sin\theta & cos\theta
\end{array}
\right )}{
\left( \begin{array}{ccc}
|^{3}P_{1} \\
|^{1}P_{1}
\end{array}
\right )}
\end{equation}

For a decay process $A\rightarrow BC$, if the initial states $A(l^{P} )$are the mixture, the partial
wave amplitude can be written as

\begin{equation}
{
\left( \begin{array}{ccc}
\mathcal{M}^{JL}_{|l+\frac{1}{2},l^{P}\rangle\rightarrow BC} \\
\mathcal{M}^{JL}_{|l-\frac{1}{2},l^{P}\rangle\rightarrow BC}
\end{array}
\right )}={
\left( \begin{array}{ccc}
cos\theta & -sin\theta \\
sin\theta & cos\theta
\end{array}
\right )}{
\left( \begin{array}{ccc}
\mathcal{M}^{JL}_{|^{3}l_{l}\rangle\rightarrow BC} \\
\mathcal{M}^{JL}_{|^{1}l_{l}\rangle\rightarrow BC}
\end{array}
\right )}
\end{equation}
In our calculations, the states $B_{1}(5721)$, $B_{s1}(5830)$ are the $1^{+}$ bottom and bottom-strange states and each of them is the mixing of $^{3}P_{1}$ and $^{1}P_{1}$ states.
In addition, we will study the strong decays of $B_{J}(5970)$ as the $2^{-}$ state and it is the mixture of $^{3}D_{2}$ and $^{1}D_{2}$ states. Considering the mixture of the
initial states, the decay width can be expressed as

\begin{equation}
\begin{aligned}
\notag
\Gamma(|l+\frac{1}{2},l^{P}\rangle\rightarrow BC)=\frac{\pi}{4}\frac{|\overrightarrow{P}|}{M_{A}^{2}}\sum_{JL}|cos\theta \mathcal{M}^{JL}_{|^{3}L_{L}\rightarrow BC}-sin\theta \mathcal{M}^{JL}_{|^{1}L_{L}\rightarrow BC}|^{2} \\
\Gamma(|l-\frac{1}{2},l^{P}\rangle\rightarrow BC)=\frac{\pi}{4}\frac{|\overrightarrow{P}|}{M_{A}^{2}}\sum_{JL}|sin\theta \mathcal{M}^{JL}_{|^{3}L_{L}\rightarrow BC}+cos\theta \mathcal{M}^{JL}_{|^{1}L_{L}\rightarrow BC}|^{2}
\end{aligned}
\end{equation}

\begin{large}
\textbf{3.1 $B_{2}^{*}(5747)$, $B_{1}(5721)$, $B^{*}_{0}$}
\end{large}
\begin{table*}[htbp]
\begin{ruledtabular}\caption{The strong decay widths of the $B_{2}^{*}(5747)$, $B_{1}(5721)$, $B^{*}_{0}$ with possible assignments. If the
corresponding decay channel is forbidden, we mak it by "-". All values in units of $MeV$.}
\begin{tabular}{c| c c c c c c}
 & \ $B_{2}^{*+}(5747)$  & \ $B_{1}^{+'}(5721)$ &\ $B_{1}^{+}(5721)$  &\ $B^{*+}_{0}$  \\
 \hline
 State &\ $1^{3}P_{2}$         &  \ $1 P_{1}'$     & \  $1P_{1}$   &\ $1 ^{3}P_{0}$ \\
 \hline
 Mass &\ $5737.2$\cite{PDG} &\  $5726.0$\cite{PDG} &\ $5726.0$\cite{PDG} &\ $5697.4$\cite{707} \\
\hline
$B^{+}\pi^{0}$ &\  $4.3$      &\  $-$   &\ $-$   &\  $76.3$   \\
$B^{*+}\pi^{0}$ &\  $3.7$      &\  $26.5$   &\ $138.8$  & \ $-$    \\
$B^{0}\pi^{+}$ &\  $8.6$     &\  $-$   &\ $-$  & \  $155.1$  \\
$B^{*0}\pi^{+}$ &\  $7.3$     &\  $13.3$   &\ $69.4$ &\  $-$    \\
total &\  $23.9$   &\  $39.8$      &\ $208.2$    &\ $231.4$ \\
\end{tabular}
\end{ruledtabular}
\end{table*}

\begin{table*}[htbp]
\begin{ruledtabular}\caption{The strong decay widths of the $B_{2}^{*}(5747)$, $B_{1}(5721)$, $B^{*}_{0}$ with possible assignments. If the
corresponding decay channel is forbidden, we mak it by "-". All values in units of $MeV$.}
\begin{tabular}{c| c c c c c c}
 & \  $B_{2}^{*0}(5747)$    &\ $B_{1}^{0'}(5721)$ &\ $B_{1}^{0}(5721)$ &\ $B^{*0}_{0}$ \\
 \hline
 State   & \ $1^{3}P_{2}$           &  \   $1 P_{1}'$ &\ $1 P_{1}$   &\ $1 ^{3}P_{0}$\\
 \hline
 Mass  &\ $5739.5$\cite{PDG}  &\ $5726.1$\cite{PDG} &\ $5726.1$\cite{PDG} &\ $5697.4$\cite{707} \\
\hline
$B^{+}\pi^{-}$    &\  $8.9$   &\   $-$   &\   $-$   &\  $78.3$  \\
$B^{*+}\pi^{-}$   &\  $7.6$   &\    $25.3$   &\   $134.9$   &\  $-$  \\
$B^{0}\pi^{0}$   &\  $4.4$   &\    $-$   &\   $-$   &\  $156.5$  \\
$B^{*0}\pi^{0}$   &\  $3.8$     &\  $12.6$   &\   $67.6$   &\  $-$  \\
total   &\  $24.7$     &\  $37.9$   &\   $202.5$   &\  $234.8$  \\
\end{tabular}
\end{ruledtabular}
\end{table*}
The bottom mesons $B_{2}^{*+}(5747)$, $B_{2}^{*0}(5747)$ are assigned to be the $2^{+}$ state with their total decay widths to be $20\pm5MeV$ and $24.2\pm1.7MeV$, respectively.
As the $1^{3}P_{2}(2^{+})$ states, we calculate their strong decay widths and the results $23.9MeV$ and $24.7MeV$ for $B_{2}^{*+}(5747)$, $B_{2}^{*0}(5747)$ are consistent well
with these experimental data. A further confirmation of this assignment is the predicted versus measured ratio of partial widths to $B^{0}\pi^{+}$ and $B^{*0}\pi^{+}$. The predicted partial ratio
\begin{equation}
\notag
\frac{\Gamma_{B_{2}^{*+}(5747)\rightarrow B^{0}\pi^{+}}}{\Gamma_{B_{2}^{*+}(5747)\rightarrow B^{*0}\pi^{+}}}=1.18
\end{equation}
is in agreement with the experimental data $1.12$, and so
does for the $B_{2}^{*0}(5747)$. As for $B_{1}^{+}(5721)$, $B_{1}^{0}(5721)$ mesons, each of them is the mixing bottom state of $^{3}P_{1}$ and $^{1}P_{1}$. In TABLE III and
TABLE IV, the $1P_{1}$, $1P_{1}^{'}$ states denote the $j_{q}=\frac{1}{2}$ and $j_{q}=\frac{3}{2}$ state, respectively. We can see that the results for $j_{q}=\frac{3}{2}(1P_{1}^{'})$
bottom states with total decay widths to be $39.8MeV$, $37.9MeV$, are roughly compatible with the experimental data $31\pm6MeV$ and $27.5\pm3.4MeV$.
These results favor $B_{1}(5721)$ to be the $j_{q}=\frac{3}{2}$ spin partner of $B_{2}^{*}(5747)$ state£¬
\begin{equation}
\notag
(B_{1}(5721),B_{2}^{*}(5747))=(1^{+},2^{+})_{\frac{3}{2}} \qquad n=1, L=1
\end{equation}
After identifying the $1P_{1}'$ assignment, the remaining $1P_{1}$ together with $1^{3}P_{0}$ state are the spin doublets with $j_{q}=\frac{1}{2}$. The
total widths of $1^{3}P_{0}$ are predicted to be $231.4MeV$, which is broader comparing with those of $j_{q}=\frac{3}{2}$ P-wave doublets. This prediction
is consistent with that of the heavy quark limit(HQL).

\begin{large}
\textbf{3.2 $B_{J}(5840)$, $B_{J}(5970)$}
\end{large}
\begin{table*}[htbp]
\begin{ruledtabular}\caption{The strong decay widths of the $B_{J}^{+}(5840)$, $B_{J}^{+}(5970)$ with possible assignments. If the
corresponding decay channel is forbidden, we mak it by "-". All values in units of $MeV$}
\begin{tabular}{c| c c| c c c c c c c c c}
 &  \multicolumn{2}{c|}{$B_{J}^{+}(5840)$}  &  \multicolumn{5}{c}{$B_{J}^{+}(5970)$} \\
\hline
States &\ $2^{1}S_{0}$ &\ $2^{3}S_{1}$ &\ $2^{3}S_{1}$ &\ $1^{3}D_{1}$ &\ $1^{3}D_{3}$ &\ $1D_{2}'$  &\  $1D_{2}$ \\
\hline
Mass  & \multicolumn{2}{c|}{$5862.9$\cite{B7}} & \multicolumn{5}{c}{$5964$\cite{PDG}}  \\
\hline
$B^{+}\pi^{0}$ &\ $-$ &\ $12.9$ &\ $10.2$ &\ $27.3 $ &\ $6.5$ &\ $-$ &\ $-$ \\
$B^{*+}\pi^{0}$ &\ $38.1$ &\ $25.4$ &\ $23.7$ &\ $14.1 $ &\ $6.0$ &\ $23$ &\ $80.9$ \\
$B^{0}\pi^{+}$ &\ $-$ &\ $25.8$ &\ $20.4$ &\ $54.6 $ &\ $13.1$ &\ $-$ &\ $-$ \\
$B^{0*}\pi^{+}$ &\ $76.1$ &\ $50.8$ &\ $47.4$ &\ $28.2$ &\ $11.9$ &\ $11.6$ &\ $40.5$ \\
$B^{+}\eta$ &\ $-$ &\ $2.7$ &\ $14.4$ &\ $25.8$ &\ $0.5$ &\ $-$ &\ $-$ \\
$B^{*+}\eta$ &\ $-$ &\ $1.6$ &\ $20.0$ &\ $8.5$ &\ $0.5$ &\ $1.2$ &\ $23.4$ \\
$B_{S}^{0}K^{+}$ &\ $-$ &\ $-$ &\ $13.1$ &\ $21.4$ &\ $0.2$ &\ $-$ &\ $-$ \\
$B_{S}^{0*}K^{+}$ &\ $-$ &\ $-$ &\ $12.3$ &\ $4.9$ &\ $0.03$ &\ $0.6$ &\ $13.9$ \\
total &\ $114.2$ & \ $121.9$ &\ $171.5$ & \ $194.3$ & \ $38.7$ &\ $36.4$ &\ $158.7$ \\
\end{tabular}
\end{ruledtabular}
\end{table*}

\begin{table*}[htbp]
\begin{ruledtabular}\caption{The strong decay widths of the $B_{J}^{0}(5840)$, $B_{J}^{0}(5970)$ with possible assignments. If the
corresponding decay channel is forbidden, we mak it by "-". All values in units of $MeV$.}
\begin{tabular}{c| c c| c c c c c c c c c}
 &  \multicolumn{2}{c|}{$B_{J}^{0}(5840)$\cite{B7}}  &  \multicolumn{5}{c}{$B_{J}^{0}(5970)$} \\
\hline
States &\ $2^{1}S_{0}$ &\ $2^{3}S_{1}$ &\ $2^{3}S_{1}$ &\ $1^{3}D_{1}$ &\ $1^{3}D_{3}$ &\ $1D_{2}'$  &\  $1D_{2}$ \\
\hline
Mass  & \multicolumn{2}{c|}{$5862.9$\cite{B7}} & \multicolumn{5}{c}{$5971.0$\cite{PDG}}  \\
\hline
$B^{+}\pi^{-}$ &\ $-$ &\ $25.8$ &\ $20.0$ &\ $54.3 $ &\ $13.4$ &\ $-$ &\ $-$ \\
$B^{*+}\pi^{-}$ &\ $76.1$ &\ $50.8$ &\ $46.7$ &\ $28.3 $ &\ $12.2$ &\ $22.9$ &\ $80.9$ \\
$B^{0}\pi^{0}$ &\ $-$ &\ $12.9$ &\ $10.0$ &\ $27.1 $ &\ $6.7$ &\ $-$ &\ $-$ \\
$B^{0*}\pi^{0}$ &\ $38.0$ &\ $25.3$ &\ $23.3$ &\ $14.1$ &\ $6.1$ &\ $11.4$ &\ $40.5$ \\
$B^{0}\eta$ &\ $-$ &\ $2.7$ &\ $14.7$ &\ $26.3$ &\ $0.5$ &\ $-$ &\ $-$ \\
$B^{*0}\eta$ &\ $-$ &\ $-$ &\ $20.9$ &\ $8.9$ &\ $0.2$ &\ $1.3$ &\ $23.9$ \\
$B_{S}^{0}K^{+}$ &\ $-$ &\ $-$ &\ $13.7$ &\ $22.1$ &\ $0.2$ &\ $-$ &\ $-$ \\
$B_{S}^{0*}K^{+}$ &\ $-$ &\ $-$ &\ $13.5$ &\ $5.2$ &\ $0.03$ &\ $0.6$ &\ $13.5$ \\
total &\ $114.1$ & \ $117.5$ &\ $162.8$ & \ $186.3$ & \ $39.3$ &\ $36.2$ &\ $158.8$ \\
\end{tabular}
\end{ruledtabular}
\end{table*}
We notice that the PDG only reported the $B_{J}(5970)$ bottom meson and omitted the $B_{J}(5840)$ state from the summary tables,
and the spin-parity of $B_{J}(5970)$ was unknown. Thus, we study the strong decay behaviors with the $2^{1}S_{0}$, $2^{3}S_{1}$
assignments for $B_{J}(5840)$ state and $2^{3}S_{1}$, $1^{3}D_{1}$, $1^{3}D_{3}$, $1D_{2}'$, $1D_{2}$ assignments for $B_{J}(5970)$
state. The results are showed in TABLE V and TABLE VI. The LHCb collaboration has suggested that the $B_{J}(5840)$, $B_{J}(5970)$
signals should be identified with the $2^{1}S_{0}$ and $2^{3}S_{1}$ bottom states. We note also that the $B\pi$ decay mode is reported by LHCb as
'possibly seen' for the strong decays of $B_{J}(5840)$ and $B_{J}(5970)$. However, our analysis indicate that the decay mode to $B\pi$ is forbidden for
$B_{J}(5840)$ as a $2^{1}S_{0}$ assignment. If the decay to $B\pi$ is confirmed in the future, the $2^{1}S_{0}$ assignment can be ruled out.
As the $2^{3}S_{1}$ assignments for $B_{J}^{+}(5840)$ and $B_{J}^{0}(5840)$, their total decay widths
are $121.9MeV$ and $117.5MeV$, and these values are compatible with the experimental data. Overall, we tentatively identify $2^{3}S_{1}$ as the assignment of
$B_{J}(5840)$.

The same with $B_{J}(5840)$, the decay channel $B\pi$ of $B_{J}(5970)$ is 'possibly seen' in experiments, so the assignments $1D_{2}^{'}$ and $1D_{2}$
are tentatively ruled out as the decay to $B\pi$ is forbidden. The experiments suggested the total decay widths for $B_{J}^{+}(5970)$ and $B_{J}^{0}(5970)$
are $62\pm20MeV$ and $81\pm12MeV$. For the assignments $1^{3}D_{3}$ and $1^{3}D_{1}$, we can see that the predicted total widths of $1^{3}D_{3}$
assignments, $38.7MeV$ and $39.3MeV$, are consistent with the experiments within the predictive power of the model and experimental uncertainties.
Thus, we slightly prefer the $1^{3}D_{3}$ assignment of the $B_{J}(5970)$. Certainly, the conclusion about the assignments
depend strongly on the accurate measurement of the decay mode $B\pi$ of $B_{J}(5840)$ and $B_{J}(5970)$.

\begin{large}
\textbf{3.3 $B_{s1}(5830)$, $B_{s2}^{*}(5840)$, $B_{s0}^{*}$}
\end{large}

\begin{table*}[htbp]
\begin{ruledtabular}\caption{The strong decay widths of the $B_{s2}^{*}(5840)$, $B_{s0}^{*}$, $B_{s1}(5830)$ with possible assignments. If the
corresponding decay channel is forbidden, we mak it by "-". All values in units of $MeV$.}
\begin{tabular}{c|c|c|cc}
 &\ $B_{s2}^{*}(5840)$  &\ $B_{s0}^{*}$  & \multicolumn{2}{c}{$B_{s1}(5830)$} \\
\hline
States &\ $1^{3}P_{2}$ &\ $1^{3}P_{0}$ &\ $1P_{1}'$ &\ $1P_{1}$  \\
\hline
Mass &\ $5839.85$\cite{PDG} &\ $5794.8$\cite{707} &\ $5828.7$\cite{PDG} &\ $5828.7$\cite{PDG} \\
\hline
$B^{+}K^{-}$ &\ $0.6$ &\ $217$ &\ $-$ &\ $-$ \\
$B^{*+}K^{-}$ &\ $0.09$ &\ $-$ &\ $1.59$ &\ $31.9$ \\
$B_{0}\overline{K^{0}}$ &\ $0.6$ &\ $217$ &\ $-$ &\ $-$ \\
$B_{0}^{*}\overline{K^{0}}$ &\ $0.06$ &\ $-$ &\ $1.51$ &\ $30.2$ \\
total &\ $1.35$ &\ $434$ &\ $3.1$ &\ $62.1$ \\
\end{tabular}
\end{ruledtabular}
\end{table*}
The bottom strange mesons $B_{s1}(5830)$ and $B_{s2}^{*}(5840)$ are identified as the $1^{+}$ and $2^{+}$ assignments in PDG,
but it is noted that the $J^{P}$ need confirmation\cite{PDG}. In order to give a further confirmation, we study the strong decay behaviors of $B_{s2}^{*}(5840)$
as the $1^{3}P_{2}$ assignment and $B_{s1}(5830)$ as the $1P_{1}'$, $1P_{1}$ assignments. The predicted total decay width of $B_{s2}^{*}(5840)$ is
$1.35MeV$ and it is consistent well with the experimental data $1.40\pm0.4$. In addition, the predicted partial decay ratio
\begin{equation}
\frac{\Gamma_{B_{s2}^{*}(5840)\rightarrow B^{*+}K^{-}}}{\Gamma_{B_{s2}^{*}(5840)\rightarrow B^{+}K^{-}}}=0.15
\end{equation}
This value is roughly compatible with the experimental data $0.093\pm0.018$, which supports $1^{3}P_{2}$
to be the assignment of $B_{s2}^{*}(5840)$. As a $1^{+}$ state, $B_{s1}(5830)$ meson is the mixture between $1^{3}P_{1}$
and $1^{1}P_{1}$. From the results in TABLE VII, we can see that the predicted total decay width of $1P_{1}'$ is $3.1MeV$
and this value is
consistent with the experimental data $0.5\pm0.4MeV$ within
the predictive power of the model. Thus, the $1P_{1}'$ is the optimal assignment for $B_{s1}(5830)$ and
we can conclude that $B_{s1}(5830)$ and $B_{s2}^{*}(5840)$ are the $j_{q}=\frac{3}{2}$ doublets,
\begin{equation}
\notag
(B_{s1}(5830),B_{s2}^{*}(5840))=(1^{+},2^{+})_{\frac{3}{2}} \qquad n=1, L=1
\end{equation}
Again, the remaining states $1P_{1}$ and $1^{3}P_{0}$ in TABLE VII are the spin doublets with $j_{q}=\frac{1}{2}$ and
their total decay widths are much broader than those of the spin doublets with $j_{q}=\frac{3}{2}$.

\begin{Large}
\textbf{4 Conclusion}
\end{Large}

In summary, we study the two-body strong decays of the excited bottom and bottom strange states
$B_{1}(5721)^{0}$, $B_{1}(5721)^{+}$,
$B_{2}^{*}(5747)^{0}$, $B_{2}^{*}(5747)^{+}$, $B_{J}(5840)^{0}$, $B_{J}(5840)^{+}$, $B_{J}(5970)^{0}$, $B_{J}(5970)^{+}$,
$B_{s1}(5830)$, $B_{s2}^{*}(5840)$ with the $^{3}P_{0}$ decay model. By analyzing the decay properties of these mesons,
we further confirm the assignments of $B_{1}(5721)$, $B_{2}^{*}(5747)$, $B_{s1}(5830)$, $B_{s2}^{*}(5840)$ and
identify the possible assignments of $B_{J}(5840)$, $B_{J}(5970)$. Our analysis support $B_{1}(5721)$ and $B_{2}^{*}(5747)$
are the spin doublets $(1^{+},2^{+})_{\frac{3}{2}}$ with $n=1$, $L=1$ and $B_{s1}(5830)$, $B_{s2}^{*}(5840)$ are the strange
partner of $B_{1}(5721)$ and $B_{2}^{*}(5747)$. The possible assignments for $B_{J}(5840)$, $B_{J}(5970)$ are $2^{3}S_{1}$
and $1^{3}D_{3}$, which need further confirmation in experiments. Especially, the decay of $B_{J}(5840)$, $B_{J}(5970)$
state to $B\pi$ is crucial to identifying the optimal assignments for these states.


\begin{large}
\textbf{Acknowledgment}
\end{large}

This work has been supported by the Fundamental Research Funds for
the Central Universities, Grant Number $2016MS133$ and Natural Science Foundation of HeBei Province, Grant Number
A2018502124.

\end{document}